\documentclass[12pt,preprint]{aastex}
\usepackage{graphicx}
\newcommand{\begen}{\begin{enumerate}}
\newcommand{\enen}{\end{enumerate}}
\newcommand{\beq}{\begin{equation}} 
\newcommand{\eeq}{\end{equation}} 
\newcommand{\beqa}{\begin{eqnarray}} 
\newcommand{\eeqa}{\end{eqnarray}}

\newcommand{\Msun}{M_{\sun}}
\newcommand{\Msunsec}{\Msun\,{\mathrm{sec}}^{-1}}

\def\ltaprx {\lower .1ex\hbox{\rlap{\raise .6ex\hbox{\hskip .3ex
        {\ifmmode{\scriptscriptstyle <}\else 
                {$\scriptscriptstyle <$}\fi}}}
        \kern -.4ex{\ifmmode{\scriptscriptstyle \sim}\else 
                {$\scriptscriptstyle\sim$}\fi}}}
\def\gtaprx {\lower .1ex\hbox{\rlap{\raise .6ex\hbox{\hskip .3ex
        {\ifmmode{\scriptscriptstyle >}\else 
                {$\scriptscriptstyle >$}\fi}}}
        \kern -.4ex{\ifmmode{\scriptscriptstyle \sim}\else 
                {$\scriptscriptstyle\sim$}\fi}}}

\begin{document}

\title{On the Contribution of Gamma Ray Bursts to the Galactic Inventory of 
Some Intermediate Mass Nuclei}
\author{Jason Pruet\altaffilmark{1}, Rebecca Surman\altaffilmark{2,3} 
\& Gail C. McLaughlin\altaffilmark{3}}
\altaffiltext{1}{N-Division, Lawrence Livermore National Laboratory,
Livermore, CA 94550; pruet1@llnl.gov}
\altaffiltext{2}{Department of Physics, Union College, Schenectady, NY 12308}
\altaffiltext{3}{Department of Physics, North Carolina State University, 
Raleigh, NC 27695-8202}

\begin{abstract}

Light curves from a growing number of Gamma Ray Bursts (GRBs) indicate that
GRBs copiously produce radioactive Ni moving outward at fractions 
of the speed of light. We calculate nuclear 
abundances of elements accompanying the outflowing Ni under the assumption 
that this Ni originates
from a wind blown off of a viscous accretion disk. We also show that GRB's
likely contribute appreciably to the galactic inventory of 
$^{42}{\rm Ca}$,
$^{45}{\rm Sc}$,
$^{46}{\rm Ti}$,
$^{49}{\rm Ti}$,
$^{63}{\rm Cu}$, 
and may be a principal site for the production of $^{64}{\rm Zn}$.

\end{abstract}

\keywords{gamma rays: bursts---nucleosynthesis---accretion disks}

\section{Introduction}

In this letter we consider the contribution of Gamma Ray Bursts to 
the Galactic inventory of some Fe-group elements. Our study is
motivated by mounting evidence that a fair fraction of GRB's are associated
with the production of a sizable amount of $^{56}{\rm Ni}$ moving
out with near-relativistic velocities \citep{pri03,sta03,hjo03,pat01,iwa98,
woo03}. The $^{56}{\rm Fe}$ to which the radioactive Ni decays is not important
for the present-day inventory of that element. As we discuss, however, 
other isotopes synthesized in the unique outflow producing $^{56}{\rm Ni}$
likely are important for Galactic chemical evolution. 

Though there are other possibilities, we will assume that GRBs are produced 
by a viscous black-hole accretion disk formed after the collapse
of a rotating massive star \citep{woo93,macfadyen}. Within the context of
this collapsar model there are two possible origins for the observed Ni. 
As in ``successful'' SNe, Ni may be synthesized explosively as a strong shock 
traverses the stellar mantle and explodes the star. \citet{mae03} have 
discussed nucleosynthetic consequences of this picture. Those authors show
that a very energetic shock driven by bi-polar jets synthesizes a peculiar 
abundance pattern that may be responsible for anomalies observed in 
extremely metal poor stars. 

The second possibility is that observed Ni comes from a vigorous
wind blown from the accretion disk \citep{mac03,macfadyen}. We discuss
implications of this scenario, which is qualitatively different from
shock-driven nucleosynthesis. Rather than develop a theory describing the
disk wind we simply begin with the assumption that a wind is responsible
for the observed Ni. As we show, this is sufficient to allow interesting
statements about the nucleosynthesis. However, it should be kept in mind
that the origin of the Ni and the nature of the GRB central engine 
remain uncertain.

\section{Nuclei Accompanying Ni in a Disk Wind}

Nucleosynthesis in the disk-wind is sensitive to the dynamic timescale
$\tau_{\rm dyn}$ characterizing the expansion of the wind, the 
entropy per baryon $s$ in the wind, and the neutron to proton ratio
$n/p$ in the wind. To a large degree these parameters are constrained
by observation. These constraints follow in part from the observations that 
Ni is ejected relativistically. Also, the efficiency for Ni production
cannot be too low. This is because the disk can eject at most one or two
solar masses of fast and initially hot material. In
order to account for the
$\sim 1/2\Msun$ of Ni inferred from the light curves of SN1998bw and SN2003dh
the mass fraction of Ni in the wind must satisfy $X({\rm ^{56}Ni})\gtrsim
1/4-1/2$ if only one or two solar masses of material is
 ejected from the disk. 

The presence of any $^{56}{\rm Ni}$ implies that $Y_e>0.485$ \citep{har85}.
A high efficiency for Ni production $(X({\rm ^{56}Ni})>0.25)$ implies
$Y_e\gtrsim 1/2$, with the exact value depending on the other outflow 
parameters. 
Detailed estimates of the electron fraction cannot be made
without reference to a specific disk/outflow model. This is because the 
temperature and density in the disk are high enough that weak interactions
determine the composition of the disk \citep{pru03,bel03,sur03} and
the composition of the wind flowing off the disk. \citet{pru203} employed
a steady-state wind picture to model the disk outflow and found 
asymptotic electron fractions anywhere in the range $0.50-0.55$, with the 
value depending sensitively on the accretion rate and viscosity of the
disk. \citet{sur03} studied the influence of charged current neutrino capture
and found that neutrino capture can increase the electron fraction 
by a considerable fraction
if the neutrino luminosity of the disk is large.  
Charged current neutrino capture is expected to 
increase the
asymptotic electron fraction in outflows from a ``canonical'' 
disk with viscosity 
$\alpha=0.1$, mass accretion rate $\dot M=0.1\Msunsec$ and Kerr parameter
$a=0.95$ by $\delta Y_e\approx 0.05$. For other trajectories/disk
models the effect may be larger. The largest $Y_e$ consistent with efficient
Ni production is $Y_e\approx 0.60$, with the exact value again depending
on other outflow parameters. 

Uncertainty in the electron fraction is somewhat of a hindrance to the
present study. However, as we show, 
overall characteristics of the nucleosynthesis 
are
relatively insensitive 
to
$Y_e$ as long as $0.505<Y_e$. The special case of $0.50<Y_e<0.505$ will
be mentioned, but not emphasized because it seems unlikely that the
electron fraction in the bulk of the outflow should be within 1\% of
the minimum value for efficient Ni production. By contrast, in
explosive burning $Y_e$ is set not by weak interactions, but by the
initial nearly isospin symmetric composition of the burning shell.

To make inferences about the dynamic timescale and entropy in the wind
we make the assumption that the wind is coasting at the low
temperatures important for nucleosynthesis. The assumption that the
wind is not accelerating at $T_9\equiv T/10^9 K < 5$ seems plausible
since an accelerating wind generally expands too quickly for efficient
Ni synthesis. Also there is little enthalpy left for driving
acceleration at such low temperatures. The great kinetic energy of the
wind and small radii at which the wind achieves low temperatures argue
that outflows from the disk are not decelerated significantly before
nuclear recombination. However, the assumption of a coasting wind will
have to be tested against numerical simulations that include 
expulsion of the stellar mantle in a consistent way.

Mass conservation determines the evolution of coasting winds through
$\dot M=4\pi r^2 \rho v_f$, with $\rho$ the density, $v_f$ the
asymptotic velocity, and $\dot M$ an effective spherical mass outflow
rate. For these winds the dynamic timescale is conveniently expressed
in terms of the entropy and 
\beq \beta\equiv { \frac{ {\dot
M}_{-1}}{v_{0.1}^3}}.  
\eeq 
Here ${\dot M}_{-1}=\dot M^{}/0.1\Msunsec$
and $v_{0.1}=v/0.1c$. To make inferences about $\beta$ and $s$, note
that efficient Ni production implies that $\beta\gtrsim 3$ if $s=50$
and $\beta\gtrsim 0.1$ if $s=30$ (see Fig. 3 in \citet{pru203}).
Also, if the Ni outflow is to be relativistic ($v_{0.1}\gtrsim1$),
$\beta$ cannot be larger than about 20.

\begin{figure}
\epsscale{0.8}
\plotone{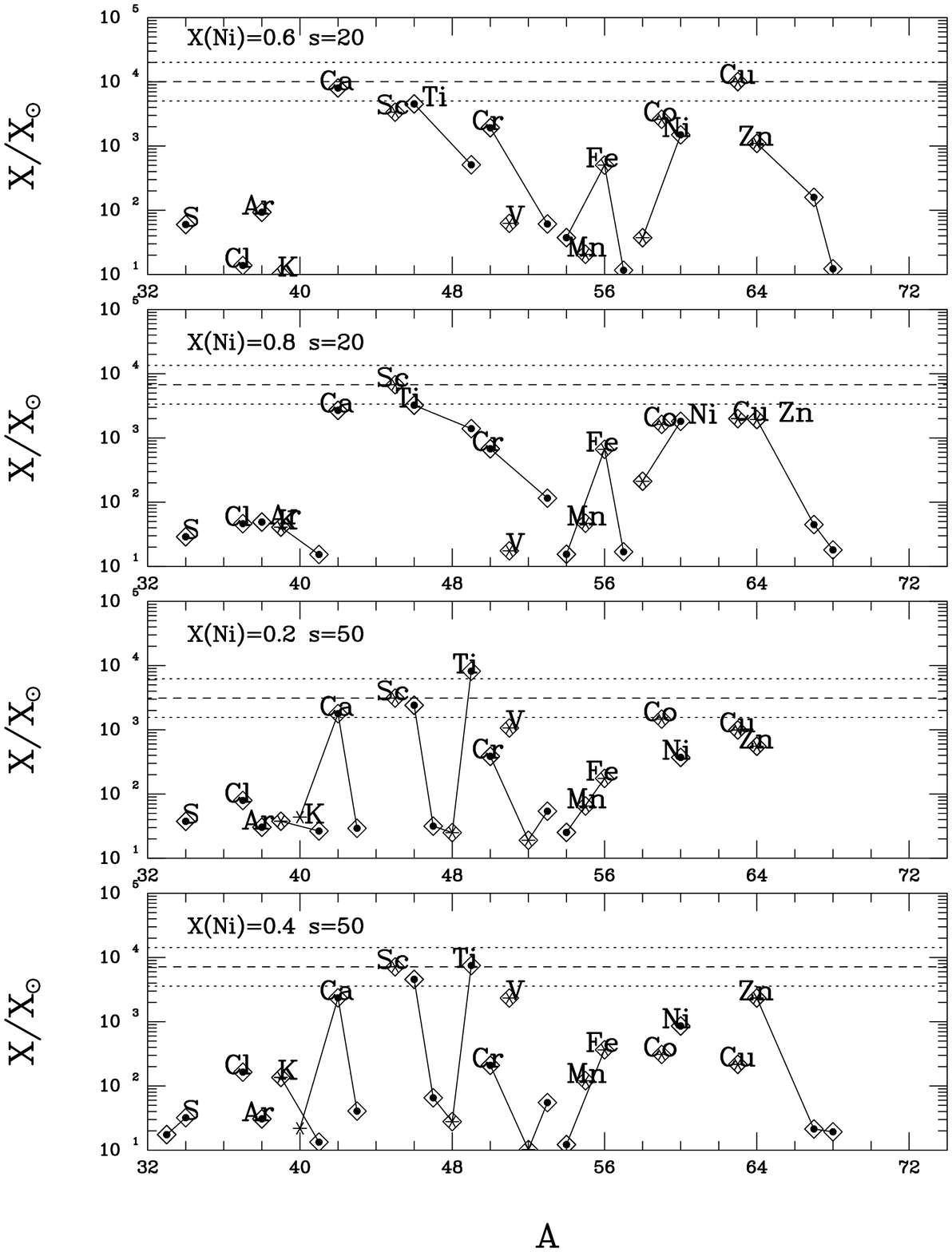}
\caption{Overproduction factors calculated for different accretion disk winds.
 Solid lines connect isotopes of a given element.
The most abundant isotope
in the Sun for a given element is plotted as an asterisk. A diamond
around a data point indicates the production of that isotope as a
radioactive progenitor.
Here $X({\rm Ni})$ is the mass fraction of $^{56}{\rm Ni}$ in the wind. Results
are shown for low entropy ($s/k_b=20$) winds that inefficiently ($\beta=0.25$)
and efficently ($\beta=16$) synthesize $^{56}{\rm Ni}$ and 
high entropy ($s=50$) winds that inefficiently ($\beta=1$) and efficiently
($\beta=16$) synthesize $^{56}{\rm Ni}$. For all cases $Y_e=0.51$. 
Note that $^{45}{\rm Sc}$, 
$^{64}{\rm Zn}$ and several other Fe-group elements have large
overproduction factors in all the wind models. \label{fig1}}
\end{figure}

With estimates for $Y_e$ and the expected range of $\beta$ and $s$, we can 
calculate nucleosynthesis in the disk wind. In Fig.~(\ref{fig1}) we show the 
results of nucleosynthesis calculations for different assumptions about the
wind parameters. These assumptions are thought to approximately bracket the
expected range of conditions in those accretion disk outflows that 
efficiently synthesize $^{56}{\rm Ni}$. The 
unnormalized overproduction factor for nucleus $j$ is defined here as
${X_j}/{X_{\sun,j}}$,
where $X_j$ is the mass fraction of the nucleus $j$ in the wind
and $X_{\odot,j}$ is the mass fraction of the nucleus in the sun.
Though the finer details of nucleosynthesis depend on the wind parameters,
there are some interesting features of the nucleosynthesis that do not.
In particular, $^{42}{\rm Ca}$,
$^{45}{\rm Sc}$,
$^{46}{\rm Ti}$,
$^{49}{\rm Ti}$,
$^{63}{\rm Cu}$, and $^{64}{\rm Zn}$ are all produced with large overproduction
factors independent of the wind parameters. 
The abundance pattern is quite different from
the abundance yields obtained in an explosive burning scenario (see Table
\ref{tbl2}). This is
not surprising, since relativistic winds and explosive shock-burning are
very different events. 

The influence of $Y_e$ on the nuclear abundances is addressed in Table
$\ref{tbl1}$, where production factors for $^{45}{\rm Sc}$ and
$^{64}{\rm Zn}$ are shown for a wind described by $\beta=4$ and
$s=30$. Here the production factor for nucleus $j$ is defined as 
\beq
O(j)=\frac{M_j}{X_{\sun,j}
M^{ej}}=\frac{M^{wind}}{M^{ej}}\frac{X_j}{X_{\sun,j}}, \eeq 
where
$M_j$ is the total mass of the nucleus $j$ ejected in the wind,
$M^{wind}\approx 1-2\Msun$ is the total mass coming from the disk in
the form of a wind, and $M^{ej}\sim 10-20\Msun$ is the total mass
ejected in the supernova explosion.  Changing $Y_e$ from $0.50$ to
$0.505$ results in dramatic changes in the nucleosynthesis. For $0.505
<Y_e<0.55$ the production factors for $^{45}{\rm Sc}$ and $^{64}{\rm
Zn}$ are within a factor of two or three of the production factors for
these nuclei at $Y_e=0.51$. For larger $Y_e$ production of $^{45}{\rm
Sc}$ approximately asymptotes to a maximimum value while production of
$^{64}{\rm Zn}$ decreases.  As mentioned above, nuclear yields are not
terribly sensitive to $Y_e$ unless weak interactions somehow conspire
to set the electron fraction within 1\% of $Y_e=0.50$. As can be seen,
an electron fraction very close to $1/2$ might have interesting
implications for $^{64}{\rm Zn}$.

For definiteness we focus in the following discussion on a
wind with $s=30$, $\beta=4$ and $Y_e=0.51$. 
These parameters are in-between the more
extreme cases shown in Fig.~(\ref{fig1}).
 In Table \ref{tbl2} we show the production 
factors for nuclei synthesized in this wind. Also
shown in Table II are production factors for the SN explosion of a 
$20\Msun$ star as calculated by \cite{woo95}. Note that the production
factors for the reference supernova are of order 10, which is the 
typical value required to explain the presence of nuclei attributed to 
type II SNe \citep{mat92}. By contrast, the production factors for the collapsar wind
are 100 or greater in some cases. To connect this with galactic chemical
evolution, note that the ``SN-equivalent production factor'' for 
collapsars is 
\beq
O^{equiv}\equiv O f_c,
\eeq
where $f_c$ is the fraction of core collapse 
SNe that become collapsars. 

The fraction of SNe that become collapsars is not well known. 
Observationally, the GRB rate is $\sim 1/100$ the SN rate 
(e.g. \citet{fra01}). This implies the rough lower limit $f_c\gtrsim 0.01$.
The true fraction is likely larger 
because special conditions must be met in order for a collapsar to
be observed as a GRB. For example, the hydrogen envelope of the progenitor star must
have been blown off in order for a jet to make its way out of the star.  
\citet{heg03} estimate that $f_c$ could be as large as $0.1$ depending
on the metallicity and initial mass function for stellar formation.

Nuclei with $O^{equiv}\sim 5-10$ will have significant contributions
from collapsars. In particular, if $f_c \gtrsim 1/50$, half or more of
the galactic abundance of $^{45}{\rm Sc}$ will come from collapsars.
If $f_c\gtrsim 0.05-0.1$ then the abundances of $^{42}{\rm Ca}$,
$^{45}{\rm Sc}$, $^{46}{\rm Ti}$, $^{49}{\rm Ti}$, $^{51}{\rm V}$,
$^{50}{\rm Cr}$, $^{63}{\rm Cu}$, and $^{64}{\rm Zn}$, can all have
significant contributions from collapsars.  A detailed
discussion of
implications of nucleosynthesis in collapsars is beyond the scope of
this Letter. Here we discuss some interesting
implications as can be inferred from the work of \citet{woo95} and
\citet{tim95}.

$^{42}{\rm Ca}$: In SNe this isotope is made during
explosive oxygen burning in sufficient quantities to explain the 
observed solar abundance. $^{40}{\rm Ca}$ is about 100 times more abundant
than $^{42}{\rm Ca}$ in SN ejecta. In the present calculations, however,
$^{42}{\rm Ca}$ and $^{40}{\rm Ca}$ are ejected with similar abundances. 
Future
observation of a large abundance of $^{42}{\rm Ca}$ relative to 
$^{40}{\rm Ca}$ in a metal-poor star might indicate enrichment by a collapsar.

$^{45}{\rm Sc}$: This is the only stable Sc isotope. Chemical evolution 
studies 
show that at low metallicities Sc is under produced by about a factor of 1.5 
relative to observation. Our calculations indicate that collapsars can
synthesize sufficient Sc to explain the discrepancy if the electron
fraction in the disk wind is larger than about $0.505$.

$^{49}{\rm Ti}$: This
isotope is under produced by about a factor of two in 
chemical evolution studies. Collapsars can explain the discrepancy if 
$f_c \sim 1/20$. 

$^{64}{\rm Zn}$: The origin of this isotope is a mystery since it  
is under produced by about a factor
of 5 in chemical evolution studies. One possible site for the origin
of this isotope is the modest entropy early-time neutrino driven 
wind occurring after core bounce in SNe \citep{woo92}. Collapsars may be 
another if the electron fraction in the disk wind is less than about $0.55$.
 
In addition to implications for chemical evolution and abundances in 
metal-poor stars, the  scenario we discuss also 
has implications for the recently observed X-ray
emission lines that have been interpreted as iron e.g.
\citep{Piro99A,Antonelli00,Piro00B}, for a review see
\citep{boettcher02}.  It has been suggested that 
these lines may instead be nickel, for example, 
either nickel produced in the disk that
lines the falls of the hole left over from the jet, or small amounts
of nickel which have been ejected at rapid velocities behind the jet 
\citep{mwbb,mw}.  As we have discussed, if the conditions are such
that $^{56}{\rm Ni}$ produced in disk winds is to account for the observed
light curves, then it is by far the most abundantly produced element
in the wind, with the second most abundant contributing at the
5\% level or less,  
either $^{60}{\rm Zn}$ or $^{58}{\rm
Zn}$.  Therefore, if
this is the mechanism for producing the observed X-ray line features,
one expects initially 
a strong line or lines from the nickel, with much weaker lines
from other elements such as zinc.  Explosive
burning on the other hand,  
produces considerable nickel, but also with
large mass fractions of light elements such as $^{24}{\rm Mg}$.

\section{Summary}

Observations of $^{56}{\rm Ni}$ associated with GRBs hint at the occurrence 
of a unique nucleosynthesis event. Massive, modest
entropy, relativistic winds may be responsible for production of observed
Ni. If so, GRBs are important for Galactic chemical evolution and can 
make important contributions to the
abundances of 
$^{42}{\rm Ca}$,
$^{45}{\rm Sc}$,
$^{46}{\rm Ti}$,
$^{49}{\rm Ti}$,
$^{63}{\rm Cu}$, and  
$^{64}{\rm Zn}$.

This work follows previous works indicating
that GRBs are associated with interesting explosive nucleosynthesis 
\citep{mae03}. As well, GRBs
may make observable amounts of deuterium \citep{pgf02,lem02,bel03},
and may be responsible for the synthesis of some r-process elements 
\citep{pru03,ino03}. Gamma Ray Bursts are emerging as important 
sources of diverse nuclei.

\acknowledgments 

JP gratefully acknowledges Stan Woosley for guidance and many
helpful suggestions.  This research has been supported through a grant 
from the DOE Program
for Scientific Discovery through Advanced Computing (SciDAC;
DE-FC02-01ER41176). This work was performed under the
auspices of the U.S. Department of Energy by the University of California,
Lawrence Livermore National Laboratory under contract W-7405-ENG-48.
GCM acknowledges support from U. S. Department of Energy under
grant DE-FG02-02ER41216.

\clearpage

\clearpage

\clearpage
\begin{deluxetable}{cccc}
\tablecaption{Sensitivity of the production of $^{64}{\rm Zn}$ and $^{45}{\rm Sc}$
to the electron fraction in the wind\label{tbl1}\tablenotemark{a}}
\tablewidth{0pt}
\tablehead{
\colhead{$Y_e$} &
\colhead{$O(^{45}{\rm Sc})$\tablenotemark{a}} &
\colhead{$O(^{64}{\rm Zn})$\tablenotemark{a}} &
}
\startdata
0.50 & 3 & 240 \\
0.505 & 270 & 125 \\
0.51 & 590 & 124 \\
0.55 & 650 &  33 \\
0.60 & 590 & 15\\


\tablenotetext{a}{Calculated assuming a total wind mass of $1\Msun$ and a 
total stellar ejecta mass of $20\Msun$.}
\enddata
\end{deluxetable}

\clearpage
\begin{deluxetable}{cccc}
\tablecaption{Production factors for a canonical wind ($\beta=4$, $Y_e=0.51$,
$s=30$) \label{tbl2}\tablenotemark{a}}
\tablewidth{0pt}
\tablehead{
\colhead{Nucleus} &
\colhead{$O$\tablenotemark{b}} &
\colhead{References $O$'s\tablenotemark{c}} &
}
\startdata
$^{42}{\rm Ca}$ & 260 & 15\\
$^{45}{\rm Sc}$ & 590 & 5\\
$^{46}{\rm Ti}$ & 211 & 11\\
$^{49}{\rm Ti}$ & 161 & 7\\
$^{51}{\rm V}$ &  61  & 7\\
$^{50}{\rm Cr}$ & 54  & 12\\
$^{53}{\rm Cr}$ & 6   & 10\\
$^{55}{\rm Mn}$ & 5   & 6\\
$^{56}{\rm Fe}$ & 23  & 5\\
$^{59}{\rm Co}$ & 73  & 7\\
$^{60}{\rm Ni}$ & 51  & 5\\
$^{63}{\rm Cu}$ & 135 & 26\\
$^{64}{\rm Zn}$ & 124 & 3\\
\tablenotetext{a}{Calculated assuming a total wind mass of $1\Msun$ and a 
total stellar ejecta mass of $20\Msun$.}
\tablenotetext{b}{Only isotopes with a production factor larger
than 5 are shown.}
\tablenotetext{c}{For model S25A of \citet{woo95}}
\enddata
\end{deluxetable}

\end{document}